\documentclass[aps,twocolumn,showpacs,pra,10pt]{revtex4-1}
\usepackage{epsfig,amsmath,color,colordvi}

\def\be{\begin{equation}}
\def\ee{\end{equation}}
\def\bee{\begin{eqnarray}}
\def\eee{\end{eqnarray}}

\def\T{{\cal T}}

\def\halb{\mbox{$\frac{1}{2}$}}

\def\with{\quad\mbox{with}\quad}

\newcommand{\bbbone}{{\mathchoice {\rm 1\mskip -4mu l}{\rm 1\mskip 
-4mu l}{\rm 1\mskip -4.5mu l}{\rm 1\mskip -5mu l}}}

\begin{document}

\title{Magnus expansion for a chirped quantum two-level system}

\author{P. Nalbach${}^{1}$ and V. Leyton${}^{2}$} 
\affiliation{${}^{1}$Westf\"alische Hochschule, M\"unsterstr. 265, 46397 Bocholt, Germany \\ 
${}^{2}$Facultad de Ciencias B\'asicas, Universidad Santiago de Cali, Calle 5 No.\ 62 -00, Santiago de Cali, Colombia}
\date{\today}
\begin{abstract}

We derive a Magnus expansion for a frequency chirped quantum two-level system. We obtain a time-independent effective Hamiltonian which generates a stroboscopic time evolution. At lowest order the according dynamics is identical to results from using a rotating wave approximation. 
We determine, furthermore, also the next higher order corrections within our expansion scheme in correspondence to the Bloch-Siegert shifts for harmonically driven systems. Importantly, our scheme can be extended to more complicated systems, i.e. even many-body systems.

\end{abstract}

\maketitle

\section{Introduction}

Driving quantum systems with external fields beyond the linear response regime has received tremendous interest in recent years. Accordingly, driven quantum many body systems exhibit, for example, nonequilibrium phase transitions in their steady state features and unusual transient decay dynamics towards the steady states \cite{Polkovnikov2011, Polkovnikov2013}. Theoretical treatment of harmonically driven quantum systems often employs a Magnus expansion \cite{Magnus54, Polkovnikov2015} to obtain an effective time-independent Hamiltonian which allows to employ the rich tool box for solving (undriven) many body problems. No such simple approach is available for more general driving.

Using tuned pulses of electric fields or lasers to steer chemical processes \cite{Prokhorenko2006} or to control quantum devices \cite{Gong2016} is another field where acurate modeling of the dynamical processes is key. Often chirped pulses are employed, as for example, for adiabatic population transfer \cite{Chelkowski1990} or to control photo-dissociation using a bond-softening mechanism \cite{Natan2012}. Recently, it has also been proposed to optimize manipulation of molecular magnets by femtosecond chirped laser pulses \cite{Zhang2012}. Furthermore, frequency-chirped drive pulses have been applied to a Josephson tunnel junction and a Josephson phase circuit \cite{Murch2010, Shalibo2012} in an effort to study the crossover from classical to quantum dynamics in these solid state qubit realizations. 

Chirping a quantum two-level system (TLS), the simplest possible quantum system, is well known and a standard procedure used already in NMR as a robust method for population inversion termed adiabatic rapid passage (ARP). Malinovsky and Krause \cite{Malinovsky2001} studied theoretically the influence of a chirped laser pulse on a quantum two-level system in great detail. Therein, they employed, as typically done, the rotating wave approximation (RWA). For harmonically driven systems the RWA coincides with the lowest order of a corresponding Magnus expansion \cite{Magnus54, Thimmel1999}, thus, justifying it for small driving strength. However, it is unclear how well the RWA performs for chirped driving or how to determine corrections as the Bloch-Siegert shift \cite{Bloch1940SiegertShift} at strong driving. Additionally, the RWA for a chirped quantum two-level system yields still a time-dependent Hamiltonian. Accordingly, for all but trivial chirped pulse protocols only numerical evaluation is possible.

In this work we devise a scheme to use the Magnus expansion for a frequency chirped TLS. We, thus, obtain a time-independent effective Hamiltonian \cite{Magnus54, Thimmel1999} which generates a stroboscopic time evolution. At lowest order the according dynamics is identical to the RWA result, thus, justifying the later. The Magnus expansion, however, allows to determine higher order effects corresponing to Bloch-Siegert shifts \cite{Bloch1940SiegertShift} known for harmonically driven systems. The presented expansion scheme can, furthermore, readily be adopted to treat chirped many body problems.

\section{Frequency Chirped TLS}

A TLS subject to a frequency chirp is described by an Hamiltonian
%
$H = \halb \Delta_0 \, \sigma_x+ \halb \epsilon_0
\, \sigma_z + \eta\,\sigma_z \sin (\omega_t \,t)$,
%
with $\hbar=1$, tunnelling amplitude $\Delta_0$, bias $\epsilon_0$, time-dependent driving frequency $\omega_t$ and driving strength $\eta$. Analytical treatment, in the regime of weak but resonant driving field $\eta\ll \Delta \simeq \omega_t$ with $\Delta=\sqrt{\Delta_0^2+\epsilon_0^2}$, is achieved after transformation in its eigenenergy basis at $t=0$,
\be
H = \halb \Delta\, \sigma_x + u\eta \, \sigma_z \sin(\omega_t\,t) + v\eta
\, \sigma_x \sin(\omega_t t)\ ,
\ee
where $u=\Delta_0/\Delta$ and $v=\epsilon_0/\Delta$. 

To proceed we switch into a reference frame rotating with the driving frequency, using the unitary transformation $R(t) = \exp\left[i\, \omega_t t \,\sigma_x/2 \right]$.
%
%
In the rotating frame the dynamics is governed by the Hamiltonian $\bar{H}(t;\omega_t) = R^\dagger (t) H(t) R(t) - R^\dagger(t) \dot{R}(t)$, i.e.
\bee \label{Hrot}
\bar{H}(t;\omega_t) &=& \halb \delta_t\, \sigma_x -\halb u\eta\, \sigma_y
+v\eta \, \sigma_x\sin(\omega_t t)
\nonumber  \\
&& + \halb u \eta\, \sigma_z \sin (2\omega_t t) + \halb u \eta\, \sigma_y \cos(2\omega_t t) \nonumber .
\eee
Herein, we denote by $\delta_t$ the detuning. For an harmonically driven TLS, i.e. a constant frequency $\omega_t=\omega_0$, the detuning is $\delta_t=\delta_0=\Delta-\omega_0$. We focus on a linear frequency chirp with chirp rate $\alpha$:
\be
\omega_t \equiv \omega(t) = \omega_0 + \alpha t\ ,
\ee
resulting in a detuning $\delta_t \equiv \delta(t) = \Delta - (\omega_0 + 2\alpha t)$.

The usually employed RWA discards all remaining oscilatory time dependent terms in (\ref{Hrot}).  Although at first glance a rather uncontrolled approximation, it coincides with the lowest order terms in the Magnus series expansion in $\eta/\omega_0$ and $\delta_0/\omega_0$ for harmonic driving \cite{Magnus54, Thimmel1999}. For a chirped TLS this justification does not hold. 

The Magnus expansion yields a time-independent effective Hamiltonian whereas the RWA Hamiltonian for a chirped TLS, $\bar{H}_{\rm   RWA} = \halb \delta(t) \sigma_x - \halb u \eta \, \sigma_y$, is still time-dependent.

The Hamiltonian $\bar{H}_{\rm RWA}$ reflects the well known Landau-Zener problem \citep{LZ32} where a quantum system is driven linearly by an external force through an avoided crossing. Herein, $u\eta$ is the coupling between the two diabatic states. The driving speed is given by $2\alpha$. The TLS is driven through its resonance by applying a chirp with square pulse profile with initial time $t_0$ and final time $t_f$ with $\omega_0+\alpha t_0 \ll \Delta \ll \omega_0 + \alpha t_f$. The Landau-Zener probability $P_{LZ}=1-\exp[-\pi u^2\eta^2/4\alpha]$ relates the probability for exciting the TLS from the ground state with the chirp rate $\alpha$. Thus, a frequency chirp allows in a very controlled way to excite the two-level system. With proper choice of the chirp rate, i.e. for $2\alpha\simeq u^2\eta^2$, well defined coherent superpositions of ground and excited states can be generated. This fact is the underlying reason for using chirped laser pulses in quantum control schemes.

\section{Magnus Expansion Scheme}

The Magnus expansion scheme \cite{Magnus54, Thimmel1999} is easily applied to harmonically driven systems but not readily usable for chirped drives. Extending the scheme for chirped TLS, we obtain a series expansion which depends additionally  on $\alpha/\omega_0^2$ and thus allows to describe {\it weak chirps}, where the frequency changes within a single period are small. 

The time evolution of a quantum system with Hamiltonian $\bar{H}(t;\omega_t)$ is described by its time evolution operator 
\be
U(t,t_0) = \T \exp\left[ i\int_{t_0}^tds\, \bar{H}(s;\omega(s)) \right] \ ,
\ee
with $\T$ being the time ordering operator. Defining the time periodic points $t_j$, for which $\omega(t_j) \, t_j = 2\pi j$, and $\bar{H}(t_j;\omega_0) = \halb \delta(t_j) \sigma_x$ and fixed $t_0=0$ we obtain
\bee
U(t,t_0) &=& \T \exp\left[i\int_{t_N}^tds\, \bar{H}(s;\omega(s))\right] \cdot
\prod_{j=1}^N U_j \with
\\ &&
U_j = \T \exp\left[ i \int_{t_{j-1}}^{t_j}ds\,
  \bar{H}(s;\omega(s)) \right] .
\eee
Therein, $N$ is choosen that $t_{N+1}>t$. In the following we employ for readability the shorthand notations
\bee
\omega_{j} \equiv \omega(t_j) &=& \omega_0+2\alpha t_{j} \\
\delta_j \equiv \delta(t_j) &=& \Delta - \omega_0-2\alpha t_j .
\eee

Defining the periods $\tau_j=t_j-t_{j-1}$, and realizing that $\bar{H}(s'+t_{j-1};\omega_0) = \bar{H}(s';\omega_{j-1})$ we find (using $s'=s-t_{j-1}$)
\be
U_j = \T \exp\left[ i \int_{0}^{\tau_j}ds\, \bar{H}(s;\omega_{j-1})\right].
\ee

Since $U_j$ is a unitary evolution, there is an effective Hamiltonian $H_{\rm eff,j}$ with $U_j=\exp\bigl[iH_{\rm eff,j}\tau_j\bigr]$. Expanding both representations of $U_j$ in a series
\bee
U_j &=& 1 + iH_{\rm eff,j}\tau_j -\halb H^2_{\rm eff,j} \tau_j^2 + 
\ldots \nonumber \\
&=& 1 + i\int_{0}^{\tau_j}ds\, \bar{H}(s;\omega_{j-1}) \nonumber \\
&& \quad + i^2 \int_{0}^{\tau_j}ds\,\int_0^{s}ds'\;
\bar{H}(s;\omega_{j-1})\bar{H}(s';\omega_{j-1}) +\ldots\, , \nonumber 
\eee
one determines $H_{\rm eff,j}$ by adding up all terms in the second line which are proportional to $i\tau_j$. The $n$-th order contribution is
\bee
I^{(j)}_n &=& \frac{i^{n-1}}{\bar{\omega}_j^n} \int_{0}^{2\pi}dz_1 
\cdots \int_0^{z_{n-1}}dz_n\; \cdot \nonumber \\
&& \qquad\quad \bar{H}(\frac{z_1}{\bar{\omega}_j};\omega_{j-1}) \cdots
\bar{H}(\frac{z_n}{\bar{\omega}_j};\omega_{j-1})\, , \nonumber 
\eee
with $\bar{\omega}_{j}$ defined by $\bar{\omega}_{j}\tau_j=2\pi$. It holds for a positive chirp, $\alpha>0$, that $\omega_{j-1}\le\bar{\omega}_j\le\omega_{j}$. Thus, we get
\be
 I^{(j)}_n = \frac{i^n}{\bar{\omega}_j^n} \cdot \sum_{k=1}^n
C^{(n,j)}_k \cdot (2\pi)^k \ee
where the coefficients $C^{(n,j)}_k$ only depend on $\eta$, $\delta_j$ and the ratio $(\omega_{j-1}/\bar{\omega}_{j})\simeq 1 -(2\pi\alpha/\omega_0^2) + O(2\pi\alpha/\omega_0^2)^2$. Thus, for $\eta, \delta_j \ll \omega_j$ and $2\pi\alpha \ll \omega_0^2$ an according series expansion can be truncated after a few terms. A tedious calculation results for the first $H^{(1)}_{\rm eff;j}$ and the second lowest order $H^{(2)}_{\rm eff;j}$ in 
\be \label{eq:eff1}
H^{(1+2)}_{\rm eff;j} = H^{(1)}_{\rm eff;j} + H^{(2)}_{\rm eff;j} 
\ee
with 
\be  \label{eq:eff2}
H^{(1)}_{\rm eff;j} = \left( \delta_{j-1} - \alpha\tau_j \right) \frac{\sigma_x}{2}  - u \eta \frac{\sigma_y}{2} 
\ee
and
\bee  \label{eq:eff3}
\hspace*{-7mm}{} H^{(2)}_{\rm eff;j} &=& -\frac{3(\eta u)^2}{4\omega_{j-1}} \frac{\sigma_x}{2} 
- u \eta \frac{\delta_{j-1}-\alpha\tau_j}{2\omega_{j-1}} \frac{\sigma_y}{2} 
+ \frac{2 \eta^2 u v}{\omega_{j-1}} \frac{\sigma_z}{2} \\
&& + \frac{4\alpha }{\omega_{j-1}^2} v \eta \frac{\sigma_x}{2} 
  +  \frac{\alpha}{\omega_{j-1}^2} u\eta \frac{\sigma_z}{2} .\label{eq:eff4}
\eee

In lowest order the dynamics during a period $\tau_j$ is governed by the mean of the detuning $\delta(t)$ during this period and the driving strength $u\eta$. The change of the former between periods is the main difference to the result for harmonic driving. The second order terms are separated in two distinct contributions. The terms of the r.h.s in Eq.(\ref{eq:eff3}) are well known from the Magnus expansion of a harmonically driven TLS. The first term corresponds to the Bloch-Siegert shift \cite{Bloch1940SiegertShift} of the resonance. The important difference here is that both, the detuning and the driving frequency, change between periods. The second order terms of Eq.(\ref{eq:eff4}) are specific to chirped TLS. These are suppressed when the change of the driving frequency within a single period is small. Note that the crossover between adiabatic to non-adiabatic driving of a Landau - Zener like dynamics is reached for chirp rates $2\alpha = u^2\eta^2$ but the Magnus expansion demands $\eta \ll \omega_t$. Thus, typically $\alpha/\omega_t^2\ll \eta/\omega_t$ except for extreme non-adiabatic chirping.

For the full time evolution operator we need besides $U_j =\exp\bigl[ iH_{\rm eff;j}^{(1)}\tau_j \bigr]$ furthermore $U(t,t_N)$.  The effective Hamiltonian (\ref{eq:eff1}) is not sufficient to describe the time evolution $U(t,t_N)$ \cite{Thimmel1999}. For simplicity, we thus restrict ourselves in the following to full periods, i.e. $U(t,t')=\prod_{j=1}^N U_j$, i.e. a stroboscopic time evolution.

\section{Stroboscopic time evolution}

The effective Hamiltonian (\ref{eq:eff1}) is time-independent within a period $\tau_j$ but, importantly, it changes between periods, thus resulting in a stroboscopic like time evolution.  

A general statistical operator $\bar{\rho}(t)=\halb(\bbbone+\vec{r}(t)\cdot\vec{\sigma})$, with the Pauli matrices $\vec{\sigma} = (\sigma_x,\sigma_y,\sigma_z)$ and their expectation values $\vec{r}(t) = (\langle\sigma_x\rangle(t),\langle\sigma_y\rangle(t),\langle\sigma_z\rangle(t))$, evolves within one period $\tau_j$ from $t_{j-1}$ to $t_j$ via $\bar{\rho}(t_{j}) = U_j^\dagger \bar{\rho}(t_{j-1}) U_j$. The expectation values follow an evolution $\vec{r}(t_j) = M_j \vec{r}(t_{j-1})$ with a the time evolution matrix
%
\bee
M_j &=& \\ 
&&\hspace*{-1cm}{\small\left( 
\begin{array}{ccc}
  c+e_x^2(1-c)     & e_xe_y(1-c)-e_zs    & e_xe_z(1-c)+e_ys \\
  e_xe_y(1-c)+e_zs & c+e_y^2(1-c)        & e_ye_z(1-c) - e_xs\\
  e_xe_z(1-c)-e_ys & e_ye_z(1-c) + e_xs  & c+e_z^2(1-c)
\end{array}
 \right) }\nonumber
\eee
%
with a representation of the Hamiltonian as $ H= \vec{e}\cdot\vec{\sigma}\, E/2$ wherein $\vec{e} = (e_x,e_y,e_z)$ is a normalized vector, and $c=\cos (E\tau)$ and $s=\sin (E\tau)$. For example, for a symmetric two-level system, i.e.  $\epsilon_0 = 0$ and, thus, $u=\Delta_0/\Delta= 1$ and $v= \epsilon_0/\Delta=0$, we find in first order $\vec{e} = (\delta_{j-1}-\alpha\tau_j,-\eta,0)/E_j$ and $E_j= \sqrt{\eta^2+(\delta_{j-1}-\alpha\tau_j)^2}$.

\section{Experimental Relevance}

Frequency-chirped drive pulses have been applied to a Josephson tunnel junction and a Josephson phase circuit \cite{Murch2010, Shalibo2012} in an effort to study the crossover from classical to quantum dynamics in these solid state qubit realizations. Employing a RWA Barth et al. \cite{Barth2011} show that frequency chirped driving of nonlinear oscillators \cite{Murch2010, Shalibo2012} causes an {\it adiabatic} evolution leading to sequential excitation of single energy levels, i.e. {\it quantum ladder climbing}. Shalibo et al. \cite{Shalibo2012} employed in their experiment a chirp from $6.1$GHz to $5.9$GHz with the resonance at about $6$GHz. Their
driving strength is $\eta=2\pi\cdot 27$MHz and the chirp rate is $\alpha=-2\pi\cdot 1$MHz/ns$=0.22\eta^2$.
%
\begin{figure}[t]
\includegraphics[width=8.5cm]{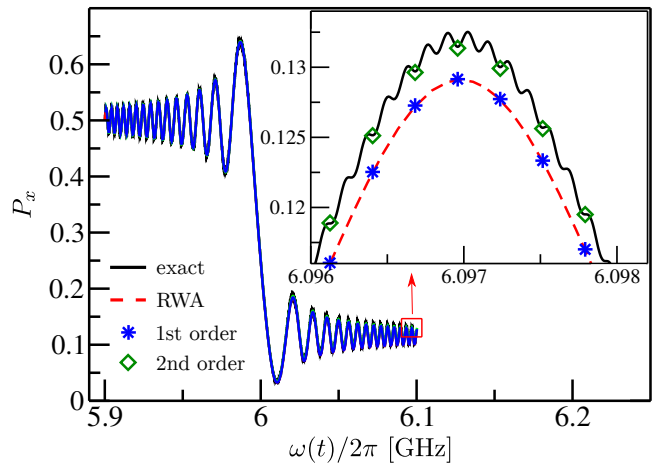}      
\caption{\label{fig3} Probability $P_x$ for adiabatic evolution versus drive frequency $\omega(t)$ at fixed bare drive period $\tau_0=0.028\eta^{-1}$ for drive parameters as in \cite{Shalibo2012}. Inside we depict a zoom to show the matching between the exact behavior (continuous line) with the second order term in the expansion (diamond symbols), and the RWA (dashed line) with the first order in the expansion (star symbols).}
\end{figure}

In Fig. \ref{fig3} we plot the expectation value $P_x=0.5\cdot (1-r_x(t))$ versus drive frequency for a chirp from $\omega(0)=2\pi\cdot 5.9$GHz to $\omega(t_N)=2\pi\cdot6.1$GHz with driving strength is $\eta=2\pi\cdot 27$MHz. We used a more non-adiabatic chirp rate $\alpha=1.5\eta^2$ and started from a coherent state, i.e. with $r_z(0)=1$. The main plot shows the basic behaviour as known for according Landau-Zener driving schemes. The inset zooms into the last oscilation during the chirp. The red dashed line shows the RWA result and the black full line the exact full dynamics which is numerically accessible for a simple TLS. Both results differ quantitatively and qualitatively. The latter shows the fast oscilatory behaviour neglected by the RWA. The blue stars reflect the stroboscopic dynamics due to the lowest order effective Hamiltonian (\ref{eq:eff2}). This lowest order dynamics is within the resolution ontop of the RWA result. The green diamonds reflect the stroboscopic dynamics due to the second order effective Hamiltonian (\ref{eq:eff1}). Due to the stroboscopic character the fast oscilatory behaviour is not resolved but quantitatively the results fall ontop of the exact result. 

Our example differs in two aspects from the experimental setting from Shalibo et al. \cite{Shalibo2012}. First we have started from a coherent state, i.e. $r_z(0)=1$ instead of from the ground state $r_x(0)=-1$ and, secondly, we used a slightly larger chirp rate pushing the chirp into the regime between adiabatic and non-adiabatic Landau-Zener dynamics. Both aspects basically increase the difference between RWA and the exact results. In detail, for the experimentally employed chirp the deviation is an order of magnitude smaller. 
Experimental chirp rates in the experiment by Shalibo et al. \cite{Shalibo2012} have been chosen which are small enough to ensure adiabaticity but large enough to minimize the needed time for the chirp in order to minimize dissipative effects \cite{Nalbach2009LZPRL, Nalbach2010LZChemPhys, Fainberg2002, Fainberg2004} due to environmental disturbances. Thus the experiment is optimally performed to see
the adiabatic evolution. However, larger chirp rates are needed to apply quantumcontrol schemes beyond pure adiabatic driving. There, the RWA might not be sufficient for accurate modelling.

\section{Conclusion}

In summary, we have studied the dynamics of a chirped two-level
system. We derived a Magnus expansion for the Hamiltonian which determines the stroboscopic dynamics of a non-harmonically driven TLS, i.e. for a linear frequency chirp. The approach is easily extended to more complicated systems, i.e. even many-body systems. 

For weak driving strength and chirp rate the lowest order effective Hamiltonian yields an identical dynamcis as the (uncontrolled) rotating wave approximation and both describe the dynamics well. For strong driving strength or large chirp rate the lowest order Magnus expansion as well as the rotating wave approximation are insufficient to model the dynamics accurately. The next higher order of the Magnus expansion, however, provides corrections which allow sufficient accuracy for the  experimentally studied case by Shalibo et al. \cite{Shalibo2012}. We propose experiments which explicitely show the breakdown of the rotating wave approximation and the relevance of the corrections which we derived.

\paragraph*{Achnowledgements}
%
PN acknowledges financial support by the DFG project NA394/2-1.

\end{document}